\documentclass[pra,aps,showpacs,twocolumn]{revtex4}
\usepackage{amssymb}
\usepackage{epsfig}
\usepackage[english]{babel}
\usepackage{latexsym}
\usepackage{graphics}
\usepackage{subfigure}

\begin{document}

\title{Manipulation of Single Neutral Atoms in Optical Lattices}
\author{Chuanwei Zhang$^{1}$, S. L. Rolston$^{2}$, and S. Das Sarma$^{1}$}

\begin{abstract}
We analyze a scheme to manipulate quantum states of neutral atoms at
individual sites of optical lattices using focused laser beams. Spatial
distributions of focused laser intensities induce position-dependent energy
shifts of hyperfine states, which, combined with microwave radiation, allow
selective manipulation of quantum states of individual target atoms. We show
that various errors in the manipulation process are suppressed below $%
10^{-4} $ with properly chosen microwave pulse sequences and laser
parameters. A similar idea is also applied to measure quantum states of
single atoms in optical lattices.
\end{abstract}

\affiliation{$^{1}$ Condensed Matter Theory Center, Department of Physics, University of
Maryland, College Park, Maryland, 20742 USA\\
$^{2}$ Department of Physics, University of Maryland, College Park,
Maryland, 20742 USA}
\maketitle

\section{Introduction}

Neutral atoms trapped in optical lattices are excellent candidates for
quantum computation because they are well isolated from environment, leading
to long coherence times, and easy to use for storing and processing quantum
information \cite{Nielsen,zoller1}. In optical lattices, controlled
interactions between atoms may be implemented effectively, and highly
entangled states of many atoms may be created in a single operational step 
\cite{Jaksch, Mandel1,Duan}. More importantly, very efficient schemes for
quantum error corrections \cite{Steane} and fault-tolerant computing \cite%
{Shor} can be straightforwardly implemented because of parallel operations
in optical lattices \cite{Briegel} and long coherence times.

An outstanding challenge for quantum computation in optical lattices is the
selective manipulation and measurement of quantum states of single atoms,
because spatial periods of typical optical lattices are shorter than optical
resolutions (long wavelength lattices may provide single site
addressability, but it is difficult to implement Mott insulating states with
one atom per site necessary for quantum computation in such lattices \cite%
{Scheun}). Single qubit operation is not only a building block of universal
quantum computation \cite{Barenco}, but also an essential ingredient for the
recently proposed one way quantum computation, where quantum information is
processed by performing single qubit rotations and measurements on entangled
cluster states \cite{Raussendorf}. Experimentally, entangled cluster states
have been realized for neutral atoms by using controlled cold collisions in
spin-dependent optical lattices \cite{Mandel1}. Therefore, implementations
of single atom manipulation and measurement may eventually lead to universal
quantum computation in optical lattices.

In this paper, we show that high fidelity selective manipulation and
measurement of single atoms in optical lattices can be achieved with the
assistance of focused lasers \cite{Rolston}. Consider a deep two dimensional
optical lattice with one atom per lattice site \cite{Greiner}. The logical
qubit basis of each atom is formed by two hyperfine states that can be
coupled coherently using microwave radiation. The couplings are same for all
atoms because of the degeneracy of their hyperfine splittings. In the
presence of spatially varying external fields, the degeneracy may be lifted
and atoms separated by certain distance can be individually manipulated. For
instance, individual atoms separated around $2.5\mu m$ have been selectively
addressed in an experiment using magnetic field gradients \cite{Schrader},
but such a method is not useful for atoms separated by a half-wavelength of
typical optical lattices, where impractically large gradients or small
fluctuations of the magnetic fields would be required.

The degeneracy of hyperfine splittings may also be lifted in the presence of
focused lasers that induce position-dependent energy shifts of hyperfine
states (\textit{light shifts}) \cite{Metcalf} due to the spatial
distributions of laser intensities. Through varying intensities and
detunings of the focused lasers, we may adjust the difference of hyperfine
splittings between neighboring atoms to a regime where microwave radiation
affects mainly target atoms, while impact on non-target atoms is strongly
suppressed with properly chosen microwave pulse sequences. Various errors in
the manipulation process are found to be below $10^{-4}$ except for the
spontaneous emission probability of atoms in the focused laser. Finally, we
show that the quantum states of individual atoms in optical lattices may
also be measured using position-dependent energy shifts of hyperfine states.

\begin{figure}[t]
\begin{center}
\resizebox*{8cm}{6cm}{\includegraphics*{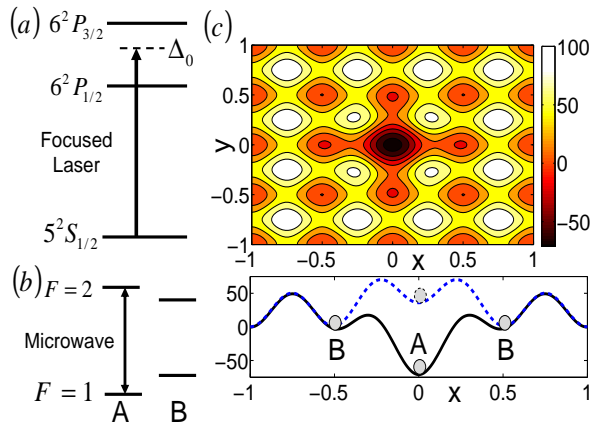}}
\end{center}
\par
\vspace*{-0.0cm}
\caption{(color online). The scheme for manipulating single atoms in optical
lattices using focused lasers and microwave radiation. (a) $^{87}$Rb
hyperfine structure. $\Delta _{0}$ is the detuning of the focused laser from
the transition $5^{2}S_{1/2}\rightarrow 6^{2}P_{3/2}$. (b)
Position-dependent hyperfine splitting induced by the spatial distribution
of the focused laser intensity. Microwave radiation couples two hyperfine
states of atoms. (c) Optical potentials for atoms around the focused laser.
The length unit is the wavelength $\protect\lambda =850nm$ of the optical
lattice. The energy unit is the recoil energy $E_{r}$. Top: a 2D plot for
atoms at state $\left\vert 0\right\rangle $. Bottom: a 1D plot along $y=0$.
Solid and dashed lines for states $\left\vert 0\right\rangle $ and $%
\left\vert 1\right\rangle $ respectively.}
\label{rr}
\end{figure}

\section{Selective Manipulation of Single Atoms}

We develop the scheme as realistically as possible, and illustrate it using $%
^{87}$Rb atoms, although the technique is applicable to other species as
well. Consider $^{87}$Rb atoms confined in a two dimensional ($xy$ plane)
optical lattice with wavelength $\lambda =850nm$. Perpendicular to the
plane, the atomic dynamics are frozen out by high frequency optical traps 
\cite{Raizen}. The lattice is ramped up adiabatically to a potential depth $%
V_{L}=50E_{r}$ such that the Bose-Einstein condensate is converted into a
Mott insulating state with one atom per lattice site \cite{Greiner}. Here $%
E_{r}$ denotes the recoil energy $E_{r}=\hbar ^{2}k^{2}/2m\approx \hbar
\times 2\pi \times 3.18KHz$. Two hyperfine ground states $\left\vert
F=1,m_{F}=-1\right\rangle \equiv \left\vert 0\right\rangle $, and $%
\left\vert F=2,m_{F}=-2\right\rangle \equiv \left\vert 1\right\rangle $ are
chosen as the logical basis of a single atom qubit at each site.

\subsection{Position-dependent hyperfine splittings}

To manipulate a target atom $A$, we adiabatically turn on a focused laser
that propagates along $\hat{z}$ axis having the maximal intensity located at 
$A$ (Fig.1). The focused laser is obtained by passing an initial large
Gaussian beam with waist $w$ through a lens with diameter $D$ and focal
length $f$, and its intensity shows an Airy pattern $I\left( r\right) =\frac{%
I\left( 0\right) }{G^{2}}\left( \int_{0}^{\frac{D}{2}}r^{\prime }J_{0}\left(
k_{f}r^{\prime }r/\sqrt{r^{2}+f^{2}}\right) \exp \left( -r^{\prime
2}/w^{2}\right) dr^{\prime }\right) ^{2}$, where $k_{f}=2\pi /\lambda _{f}$
is the wavevector of the focused laser, $J_{0}\left( r\right) $ is the zero
order Bessel function, and $G=\int_{0}^{D/2}r^{\prime }J_{0}\left( 0\right)
\exp \left( -r^{\prime 2}/w^{2}\right) dr^{\prime }$. With properly chosen
lens parameters (see Tab I), the initial Gaussian beam is focused to the
diffraction limit, where the intensity pattern is accurately described by
above Airy function. In contrast, commonly used Gaussian beam approximation
is not precise in such limit. In addition, the Airy pattern has a narrower
waist for the central intensity distribution, leading to a smaller focused
laser intensity on neighboring atoms, although the pattern extends more
broadly than a Gaussian distribution that has an exponential decay.

Such an intensity pattern induces position-dependent energy shifts \cite%
{Jessen}%
\begin{equation}
\Delta E_{i}\left( r\right) =\frac{3\pi c^{2}}{2}I\left( r\right)
\sum_{j\left( \neq i\right) }\frac{\Gamma _{j}\left\vert c_{ij}\right\vert
^{2}}{\omega _{ij}^{3}\Delta _{ij}}  \label{eshift1}
\end{equation}%
for hyperfine ground states $\left\vert 0\right\rangle $ and $\left\vert
1\right\rangle $, where $c$ is the speed of light, $\Gamma _{j}$ is the
decay rate of excited state $\left\vert j\right\rangle $, $c_{ij}$ is the
transition coefficient, $\omega _{ij}$ is the frequency, and $\Delta _{ij}$
is the detuning of the focused laser for the transition $\left\vert
i\right\rangle \rightarrow \left\vert j\right\rangle $. Different
polarizations and detunings of the focused laser yield different shifts of
hyperfine splittings $\left\vert \Delta E\left( r\right) \right\vert
=\left\vert \Delta E_{1}\left( r\right) -\Delta E_{0}\left( r\right)
\right\vert $ between two qubit states. Here we choose a $\sigma ^{+}$%
-polarized laser that drives the $5S\rightarrow 6P$ transition \cite%
{footnote} to obtain a small diffraction limit (Fig.1(a)). The laser induces
a red-detuned trap for state $\left\vert 0\right\rangle $, but a
blue-detuned trap for state $\left\vert 1\right\rangle $. The wavelength $%
\lambda _{f}\approx 421nm$ (i.e. the detuning $\Delta _{0}$) is optimized to
obtain the maximal ratio between energy splitting of two qubit states and
the spontaneous scattering rate \cite{Mandel1}.

\begin{table}[b]
\caption{Experimental parameters. $\protect\delta \left( \protect\lambda %
/2\right) $ is the difference of the hyperfine splittings between atoms $A$
and $B$. For other atoms, $\protect\delta $ is slightly larger. $\protect%
\tau $ is the spontaneous emission probability for atoms in the focused
laser during the single qubit manipulation process. $P_{f}$ is the power of
the focused laser. The recoil energy $E_{r}=\hbar \times 2\protect\pi \times
3.18KHz$.}%
\begin{ruledtabular}
\begin{tabular}{cccccccccccc}
$\lambda $ & $V_{L}$ & $D$ & $f$ & $w$ & $\lambda _{f}$ & $\frac{\Delta _{0}}{2\pi
\hbar} $   
  \\ 
\hline 
850$nm$ & 50$E_{r}$ & 20$mm$ & 20$mm$ & 20$mm$ & 421$nm$ & -1209$GHz$ &  \\
\hline
\hline
$|\Delta E\left( 0\right)| $ & $\hbar \delta \left( \frac{\lambda }{2}\right) $ & $\hbar \omega_{0}$ & $\Delta V$ & $\tau $ & $P_{f}$ \\
\hline
107$E_{r}$ & 102$E_{r}$ & 12.8$E_{r}$ & 15$E_{r}$ & 6$\times 10^{-4}$ & 17$\mu$W

\end{tabular}
\end{ruledtabular}
\end{table}

Because of the inhomogeneity of the focused laser intensity, $\left\vert
\Delta E\left( r\right) \right\vert $ reaches a maximum at the target atom $%
A $ and decreases dramatically at neighboring sites, as shown in Fig. 1.
Therefore the degeneracy of hyperfine splittings between different atoms is
lifted (Fig.1(b)). In Fig.1(c), the optical potential $V$ for atoms around
the focused laser is plotted. We see that the minimal potential barrier, $%
\Delta V\approx 20E_{R}$, occurs for atom $B$ at state $\left\vert
0\right\rangle $. The tunneling rate of atom $B$ to the neighboring site $A$
is $\varpi \approx 2\pi J^{2}/\hbar E_{g}$ using Fermi's Golden rule, where $%
J$ is the hopping matrix element and $E_{g}$ is the energy gap between the
initial and final states. In a symmetric optical lattice, $E_{g}$ is just
the on-site interaction between two atoms. The asymmetry of sites $A$ and $B$
yields an energy gap $E_{g}$ on the order of the trapping frequency of site $%
A$, which strongly suppresses the tunneling rate. A rough estimate shows
that the tunneling time $1/\varpi $ is about $13s$ for the parameters in
Tab. I.

To avoid excitations of atoms to higher bands of optical lattices, the
rising speed of the focused laser intensity should satisfy the adiabatic
condition $\hbar \left\vert d\Delta E_{i}\left( 0\right) /dt\right\vert =\xi
\left\vert \omega _{eg}\right\vert ^{2}/\left\vert \left\langle \Phi
_{e}\right\vert \partial V/\partial \Delta E_{i}\left( 0\right) \left\vert
\Phi _{g}\right\rangle \right\vert $, where $\omega _{eg}$ is the energy
gap, $\left\vert \Phi _{g}\right\rangle $ and $\left\vert \Phi
_{e}\right\rangle $ are the wavefunctions of the ground and excited states,
and $\xi \ll 1$ is a parameter that determines the degree of adiabaticity.
Because of the high potential barrier for each atom, the wavefunctions and
energy gap may be obtained using the harmonic oscillator approximation, and
we find $\left\vert \frac{d\Delta E_{i}\left( 0\right) /E_{r}}{dt}%
\right\vert =\xi \frac{8\bar{w}^{2}E_{r}\left( \Delta V_{i}\left( r\right)
/E_{r}\right) ^{5/4}}{\hbar a_{0}r\exp \left( -2r^{2}/\bar{w}^{2}\right) }$,
where $a_{0}=\sqrt{\hbar ^{2}/mE_{r}}$, $\Delta V_{i}\left( r\right) $ is
the potential barrier for atom at position $r$ with state $\left\vert
i\right\rangle $, and we have approximated the Airy pattern of the focused
laser intensity using a Gaussian function with an effective beam waist $\bar{%
w}$. The rising speed is limited by atom $B$ at state $\left\vert
0\right\rangle $ that has minimal potential barrier, and is estimated to be $%
\left\vert \frac{d\Delta E_{0}\left( 0\right) /E_{r}}{dt}\right\vert \approx
3.6\times 10^{6}\xi \left( \Delta V\left( \lambda /2\right) /E_{r}\right)
^{5/4}$Hz. The total ramping up time is found to be $57\mu s$ for $\xi
=0.005 $, which corresponds to a $10^{-4}$ probability for excitation to
higher bands.

\subsection{Single qubit rotation}

The position-dependent hyperfine splittings induced by the focused laser,
combined with microwave radiation, can be used to perform arbitrary single
qubit unitary operations on the target atom $A$. The microwave frequency is
chosen to be resonant with the hyperfine splitting of $A$, and has a $\delta
\left( r\right) =\left( \left\vert \Delta E\left( 0\right) \right\vert
-\left\vert \Delta E\left( r\right) \right\vert \right) /\hbar $ detuning
for non-target atoms (Fig. 1(b)). The coupling between two states $%
\left\vert 0\right\rangle $ and $\left\vert 1\right\rangle $ is described by
the Rabi equation%
\begin{equation}
i\frac{d}{dt}\left( 
\begin{array}{c}
c_{0} \\ 
c_{1}%
\end{array}%
\right) =\left( 
\begin{array}{cc}
0 & e^{-i\chi }\Omega \left( t\right) /2 \\ 
e^{i\chi }\Omega \left( t\right) /2 & -\delta%
\end{array}%
\right) \left( 
\begin{array}{c}
c_{0} \\ 
c_{1}%
\end{array}%
\right) ,  \label{rabieq}
\end{equation}%
where $\Omega $ is the Rabi frequency and $\chi $ is the phase of the
microwave. The evolution of the quantum states is equivalent to the rotation
of a spin with components $S_{z}=\left\vert c_{1}\right\vert ^{2}-\left\vert
c_{0}\right\vert ^{2}$, $S_{x}=2\cos \theta \left\vert c_{1}c_{0}\right\vert 
$, $S_{y}=2\sin \theta \left\vert c_{1}c_{0}\right\vert $ on a Bloch sphere 
\cite{Metcalf}, where $\theta $ is the relative phase between two states. $%
\delta $ is the rotation frequency along $\hat{S}_{z}$ axis, while $\Omega
\left( t\right) $ is the frequency along an axis on $xy$ plane whose
direction is determined by phase $\chi $. For instance, $\chi =0$ and $\pi
/2 $ correspond to rotations along axes $\hat{S}_{x}$ and $\hat{S}_{y}$
respectively, and the combination of them may implement arbitrary single
qubit unitary operations \cite{Nielsen}. In this paper we focus on $\chi =0$
although similar results for $\chi =\pi /2$ may be obtained
straightforwardly.

\begin{figure}[t]
\begin{center}
\resizebox*{8cm}{4cm}{\includegraphics*{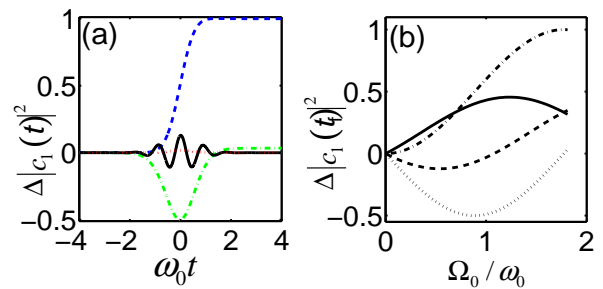}}
\end{center}
\par
\vspace*{-0.0cm}
\caption{(color online). (a) Time evolution of $\Delta |c_{1}(t)|^{2}$
during the microwave pulse. $\Omega _{0}/\protect\omega _{0}=1.81$ is the
Rabi amplitude for a pulse whose area is slightly larger than $\protect\pi $%
. Atoms $A$ (dashed dotted) and $B$ (solid) with initial state $\left(
c_{0},c_{1}\right) =\left( \protect\sqrt{\frac{1}{2}},i\protect\sqrt{\frac{1%
}{2}}\right) $; atoms $A$ (dashed) and $B$ (dotted) with initial state $%
\left( 1,0\right) $. (b) $\Delta |c_{1}(t_{f})|^{2}$ is the population
variation of atom $A$ after the microwave pulse. Initial states $\left(
1,0\right) $ (dashed dotted); $\left( \protect\sqrt{\frac{1}{2}},i\protect%
\sqrt{\frac{1}{2}}\right) $ (dotted), $\left( \protect\sqrt{\frac{2}{3}},%
\protect\sqrt{\frac{1}{4}}+i\protect\sqrt{\frac{1}{12}}\right) $ (dashed),
and $\left( \protect\sqrt{\frac{2}{3}},\protect\sqrt{\frac{1}{4}}-i\protect%
\sqrt{\frac{1}{12}}\right) $ (solid).}
\end{figure}

To achieve a reliable single qubit rotation with negligible impact on
non-target atoms, the microwave Rabi pulses need to satisfy the following
criteria: (i) the variations of the occupation probabilities of state $%
\left\vert 1\right\rangle $ of non-target atoms are extremely small (below $%
10^{-4}$); (ii) the relative phase $\theta $ between two qubit states must
return to the initial value after the microwave pulses. In the following, we
show how a single qubit rotation on the target atom $A$ satisfying these
criteria is achieved by using Gaussian shaped rotation pulses together with
refocusing pulses.

A Gaussian shaped pulse $\Omega \left( t\right) =\Omega _{0}\exp \left(
-\omega _{0}^{2}t^{2}\right) $ \ ($-t_{f}\leq t\leq t_{f}$) is used to
perform a single qubit rotation on the target atom $A$. In the frequency
domain, the Fourier transformation $\Omega \left( \omega \right) $ of the
Rabi frequency $\Omega \left( t\right) $ of this type of pulse shows a
Gaussian shaped decay with respect to the detuning $\delta $ from the
microwave frequency. Because of their large detuning $\delta $, non-target
atoms undergo Rabi oscillations with small frequencies $\Omega \left( \delta
\right) $, which strongly suppress the variations of their occupation
probabilities.

This scenario is confirmed by numerically integrating Eq. (\ref{rabieq})
with different initial states and calculating the variation $\Delta
|c_{1}(t)|^{2}=|c_{1}(t)|^{2}-|c_{1}(-t_{f})|^{2}$ of the occupation
probabilities at state $\left\vert 1\right\rangle $. A small constant $%
\omega _{0}=\frac{1}{8}\delta \left( \frac{\lambda }{2}\right) $ for the
microwave is chosen to avoid large impact on non-target atoms, and different
pulse areas are implemented by varying the pulse amplitude $\Omega _{0}$,
instead of the pulse period $2t_{f}$. In Fig. 2(a), $\Delta |c_{1}(t)|^{2}$
of atoms $A$ and $B$ are plotted with respect to the scaled time $\omega
_{0}t$. We see large variations of $\left\vert c_{1}\right\vert ^{2}$ for
atom $A$ after the pulse, which correspond to a single qubit rotation. In
comparison, there are no obvious changes of $\left\vert c_{1}\right\vert
^{2} $ for the neighboring atom $B$. In Fig. 2(b), the total changes of $%
\left\vert c_{1}\right\vert ^{2}$ for the target atom $A$ at time $t_{f}$
are plotted with respect to $\Omega _{0}/\omega _{0}$. We see different
pulse areas can be achieved by adjusting $\Omega _{0}$.

For the target atom\ ($\delta =0$), the time evolution of $\left\vert
c_{1}\left( t\right) \right\vert ^{2}\,$can be obtained analytically by
solving Eq. (\ref{rabieq}), which yields%
\begin{equation}
\left\vert c_{1}\left( t\right) \right\vert ^{2}=\left( 1-\rho \sin \left(
\eta \Omega _{0}/\omega _{0}+\phi \right) \right) /2.  \label{ana}
\end{equation}%
Here $\rho =\sqrt{1-4\left\vert c_{0}\left( -t_{f}\right) \right\vert
^{2}\left\vert c_{1}\left( -t_{f}\right) \right\vert ^{2}\cos ^{2}\theta
\left( -t_{f}\right) }$, $\eta =\omega _{0}\int_{-t_{f}}^{t}\exp \left(
-\omega _{0}^{2}t^{\prime 2}\right) dt^{\prime }$ and $\phi =\arcsin \left(
\left( 1-2\left\vert c_{1}\left( -t_{f}\right) \right\vert ^{2}\right) /\rho
\right) $. This expression is in very good agreement with the numerical
results presented in Fig. 2. 
\begin{figure}[t]
\begin{center}
\vspace*{-0.0cm}
\par
\resizebox *{8cm}{5cm}{\includegraphics*{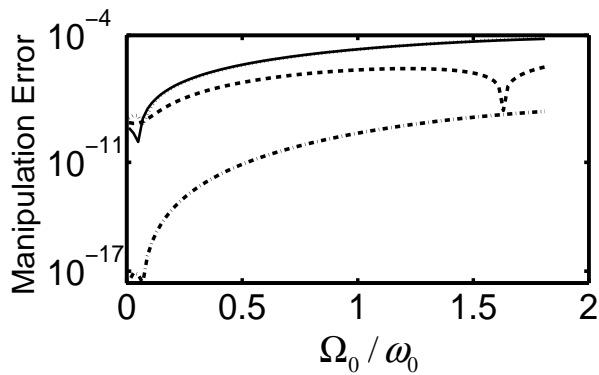}}
\end{center}
\par
\vspace*{-0.5cm}
\caption{Manipulation error $\protect\epsilon =\left\vert \left\vert
c_{1}\left( t_{f}\right) \right\vert ^{2}-\left\vert c_{1}\left(
-t_{f}\right) \right\vert ^{2}\right\vert $ for the nearest neighboring atom 
$B$. $\protect\epsilon $ is smaller for other atoms because they have larger
detuning $\protect\delta \left( r\right) $. $-7\leq \protect\omega _{0}t\leq
7$ in the numerical integration of Eq. (\protect\ref{rabieq}). Initial
states are same as those in Fig. 2(b).}
\label{rr3}
\end{figure}
In Fig. 3, we plot the single qubit manipulation errors for the nearest
neighboring atom $B$ with respect to $\Omega _{0}/\omega _{0}$. The errors
are smaller than $10^{-4}$ for different Rabi pulse areas and different
initial states. We have also performed numerical simulations for many other
initial states, finding that errors below $10^{-4}$ are always achieved. The
manipulation errors are even smaller for all other atoms because they have
larger detuning $\delta \left( r\right) $.

The relative phase $\theta $ is harder to control than the occupation
probability because it can not only vary from $0$ to $2\pi $ many times with
large frequency $\delta \left( r\right) $ during the Rabi pulses, but also
be easily affected by interactions with the environment that lead to
dephasing. To eliminate the variation of $\theta $, we use the following
pulse sequence that is similar to the refocusing process in NMR studies \cite%
{Nielsen}. To realize an area $\alpha $ pulse, we adiabatically ramp up the
focused laser, apply a $\alpha /2$ pulse on the target atom $A$ (it induces
a rotation around $\hat{S}_{x}$ axis) using the Gaussian shaped pulse
described above, then adiabatically ramp down the focused laser in step (i).
In this step, atoms obtain phase variations $\Delta \theta $ determined by
their detunings $\delta \left( r\right) $ and the dephasing process. In step
(ii), we apply a fast resonant $\pi $ pulse to all atoms that corresponds to
an angle $\pi $ rotation around $S_{x}$ axis. This pulse is called the
refocusing pulse. In step (iii) we repeat step (i), and in step (iv) we
apply another $\pi $ refocusing pulse. The two refocusing pulses do not
affect the pulse area for target atoms because the combination of them
corresponds to a $2\pi $ rotation around $S_{x}$ axis. However, the
effective directions of the rotations around $S_{z}$ axis for all atoms in
step (iii) are reversed from that in step (i) by two refocusing pulses,
therefore the phase variation in step (iii) becomes $-\Delta \theta $. These
two phase variations cancel each other and the relative phase $\theta $
returns to its initial value after the pulse sequence.

\subsection{Fidelity analysis}

In an experiment, a difficult parameter to specify is the spatial
distribution of the focused laser intensity that determines the detuning $%
\delta \left( r\right) $, the central parameter of the scheme. However, our
scheme only requires a large $\delta \left( r\right) $, not any specific
value, therefore this is not a significant difficulty. Another important
issue in the experiment is the misalignment of the focused laser from the
minimum of the optical lattice potential. A small displacement $\Delta x$ of
the focused laser induces a detuning of the microwave scaling as $\left(
\Delta x\right) ^{2}$ from the hyperfine splitting between two qubit states
of the target atom, and thus reduces the fidelity of the single qubit
rotation. For $\Delta x=1nm$, we estimate the detuning to be $2\pi \times
3Hz $ and the fidelity of the single qubit rotation is degraded by $2\times
10^{-4}$. Generally, errors due to small mis-detuning of the microwave may
be corrected using composite pulses technology developed in\ the NMR quantum
computation \cite{Jones}.

In the single qubit manipulation, the probability of spontaneous scattering
one photon for atoms in the focused laser is estimated to be $\tau =6\times
10^{-4}$, which is the worst parameter in the scheme. This parameter is
limited by the need for a maximal ratio between the vector light shifts and
the spontaneous scattering rate, which does not allow arbitrarily large
detunings.

\section{Selective Detection of Single Atoms}

The energy shift induced by the focused laser is around several hundreds $%
KHz $, which is much smaller than the decay rate $\Gamma $ of excited states
and cannot be used to selectively measure quantum states of target atoms. To
obtain large energy shifts, we transfer the target atom to other magnetic
sublevels ($\left\vert 0\right\rangle \rightarrow \left\vert \bar{1}%
\right\rangle \equiv \left\vert F=2,m_{F}=2\right\rangle $, $\left\vert
1\right\rangle \rightarrow \left\vert 0\right\rangle $) using several $\pi $
microwave pulses with the assistance of focused lasers. Then a focused $%
\sigma ^{+}$-polarized laser resonant with the transition $\left\vert \bar{1}%
\right\rangle \rightarrow $ $\left\vert 6^{2}P_{3/2}:F^{\prime
}=3,m_{F^{\prime }}^{\prime }=3\right\rangle $ is applied to detect atoms at
state $\left\vert \bar{1}\right\rangle $. In this process, resonant
fluorescence is observed if and only if the initial state of the target atom
is $\left\vert 0\right\rangle $ and the probability to detect an atom yields 
$\left\vert c_{0}\right\vert ^{2}$ of the target atom. The detection laser
can be focused on the target atoms as well as the focused laser beam using
for single atom manipulation because they both use $5S\rightarrow 6P$
transition. In the experiment, a constant $30G$ magnetic field along $x$%
-axis may be used to induce about $2\pi \times 84MHz$ Zeeman splitting
between states $\left\vert 1\right\rangle $ and $\left\vert \bar{1}%
\right\rangle $, while $2\pi \times 111.6MHz$ between $\left\vert
6^{2}P_{3/2}:F^{\prime }=3,m_{F^{\prime }}^{\prime }=-1\right\rangle $ and $%
\left\vert 6^{2}P_{3/2}:F^{\prime }=3,m_{F^{\prime }}^{\prime
}=3\right\rangle $ \cite{Metcalf}. Therefore the detection laser is about $%
\delta _{1}=2\pi \times 27.6MHz$ detuned from the transition $\left\vert
1\right\rangle \rightarrow $ $\left\vert 6^{2}P_{3/2}:F^{\prime
}=3,m_{F^{\prime }}^{\prime }=-1\right\rangle $, which yields a maximal
ratio $\left( \Gamma _{1}/2\delta _{1}\right) ^{2}I\left( r\right) /I\left(
0\right) \lesssim 2\times 10^{-5}$ \cite{Metcalf} between the photons
scattering on non-target and target atoms, that is, $5\times 10^{4}$ photons
have been scattered on the target atoms to induce one photon scattering on
the neighboring non-target atoms. The impact on non-target atoms in the
detection process may therefore be neglected.

\section{Discussion and Conclusion}

We emphasize that such high fidelity single atom manipulation cannot be
accomplished using magnetic field gradients. To obtain the same detuning $%
\delta \left( r\right) $ as that using focused lasers, an impractical
magnetic field gradient\ ($\sim 3.6$T/cm) is required. While in a typical
gradient $10$G/cm, a single atom rotation takes about $40$ms (much longer
than the cold collision time ($\sim 100\mu s$) for many-qubit gates \cite%
{Jaksch,Mandel1}), and requires unattainably small magnetic field
fluctuations ($\sim 10^{-6}$G). In addition, the magnetic field gradient
method can only be used to selectively address atoms along one dimension,
while the focused laser scheme can select two dimensions.

We notice that single qubit rotations may also be performed using two-photon
Raman transitions \cite{Yavuz} with the assistance of focused lasers. In
this case, the beam waists of the Raman pulses need to be relatively large ($%
\sim 2\mu m$) so that small misalignments of the lasers do not cause large
changes of the laser intensities on the target atoms that may diminish the
fidelity of single qubit rotations. Because two-photon Raman transitions
only affect atoms inside the Rabi pulses, atoms separated by a long distance
may be addressed simultaneously with different pulse sequences, which may
reduce the total computation time.

In summary, we have analyzed a scheme for manipulating and measuring quantum
states of single atoms in optical lattices with the assistance of focused
lasers. With properly chosen experimental parameters, various manipulation
errors are suppressed below $10^{-4}$ except the spontaneous emission
probability of atoms in focused lasers. Our proposal includes a realistic
and practical quantitative analysis suggesting plausible experimental
implementation of single atom manipulation in optical lattice quantum
computation architecture. The technique also has broad applications for
investigating various interesting physics in optical lattices such as
engineering and probing many-body quantum states in strongly-correlated
systems, topological quantum computation, etc. \cite{Zhang}. We believe that
our work establishes a practical and workable method for single atom
manipulation in optical lattices.

\begin{acknowledgments}
We thank V. Scarola for valuable discussion. This work is supported by
ARO-DTO, ARO-LPS, and NSF.
\end{acknowledgments}


\begin{references}
\bibitem{footnote} Only transitions to $6P$ states are considered because the detunings to
other states are much larger.
\end{references}
\end{document}